\documentstyle[aps,eqsecnum,epsf,feynmp,
]{revtex}
\newcommand{\onabla}{\overline{\nabla}}
\newcommand{\means}[1]{\left\langle #1 \right\rangle_\mu^{\rm s}}
\newcommand{\cums}[1]{\left\langle #1 \right\rangle_{\mu,c}^{\rm s}}
\newcommand{\meanm}[1]{\left\langle #1 \right\rangle_\mu}

\def\setval{\fmfset{wiggly_len}{2mm}\fmfset{arrow_len}{2mm}\fmfset{dash_len}{1.5mm}
\fmfpen{0.125mm}\fmfset{dot_size}{1thick}}
\begin{document}
\title{
Fluctuation Pressure of a Stack of Membranes}
\author{M. Bachmann\thanks{email: mbach@physik.fu-berlin.de},
H. Kleinert\thanks{email: kleinert@physik.fu-berlin.de}, and
A. Pelster\thanks{email: pelster@physik.fu-berlin.de}}
\address{Institut f\"ur Theoretische Physik, Freie Universit\"at Berlin,
Arnimallee 14, 14195 Berlin, Germany}
\date{\today}
\maketitle
\begin{abstract}
We calculate the universal pressure constants of a stack of $N$ membranes between walls
by strong-coupling theory. The
results are in very good agreement with values from Monte-Carlo simulations. 
\end{abstract}
\section{Introduction}
Membranes formed by lipid bilayers are important
biophysical systems occuring as boundaries of organells and vesicles. 
In equilibrium with a reservoir of molecules,
tension vanishes and 
the shape is governed by extrinsic curvature energy $E_C$.
If a stack of membranes is placed between two parallel walls, violent thermal out-of-plane 
fluctuations of the membranes exert a pressure $p$ upon the walls. The pressure law 
was found by Helfrich~\cite{helf1} and reads for $N$ membranes
\begin{equation}
\label{press}
p_N=\frac{2N}{N+1}\alpha_N \frac{(k_BT)^2}{\kappa a^3},
\end{equation}
where $L=(N+1)a$ is the distance between the walls, and $\kappa$ the bending stiffness. 
The universal pressure
constants $\alpha_N$ are not calculable exactly. For a single membrane, $\alpha_1$ was 
roughly estimated by theoretical~\cite{helf1} and 
Monte-Carlo methods~\cite{jk,jkm,janke,GK}.
The most precise theoretical result was obtained by 
strong-coupling theory~\cite{mem}
yielding $\alpha_1^{\rm th}=0.0797149$ which lies well within the 
error bounds of the latest Monte-Carlo estimate 
$\alpha_1^{\rm MC}=0.0798\pm 0.0003$~\cite{GK}. 

For more than one membrane between the walls, the strong-coupling calculation of 
Ref.~\cite{mem} must be modified in a nontrivial way. We must find a different potential
which keeps the membranes apart and whose strong-coupling limit ensures non-interpenetration. 
For this we take advantage of the fact that
membranes between walls have similar properties as a stack of nearly parallel
strings fluctuating in a plane
between line-like 
walls~\cite{lip1,lip2}, in particular the same type of pressure law (\ref{press}) with 
$\kappa$ substituted by the string tension $\sigma$. 
The characteristic universal constants
of the latter system were {\it exactly} calculated in Refs.~\cite{GK,lip1} from an analogy 
to a gas of fermions
in $1+1$ dimensions~\cite{deGennes,pok,thacker}. We use these exact values to determine
a potential which, when applied to the stack of membranes, yields a perturbation
expansion for the pressure
constants for an {\it arbitrary} number of membranes to be evaluated in the 
strong-coupling limit of complete repulsion. 

Our results are in excellent agreement with all available Monte-Carlo
estimates~\cite{jkm,janke,GK} for $N=1,3,5$. 
By an extrapolation to $N\to \infty$ we determine the pressure constant $\alpha_\infty$ 
for infinitely many membranes.
\section{Stack of Strings}
\begin{figure}
\centerline{\epsfxsize=12cm \epsfbox{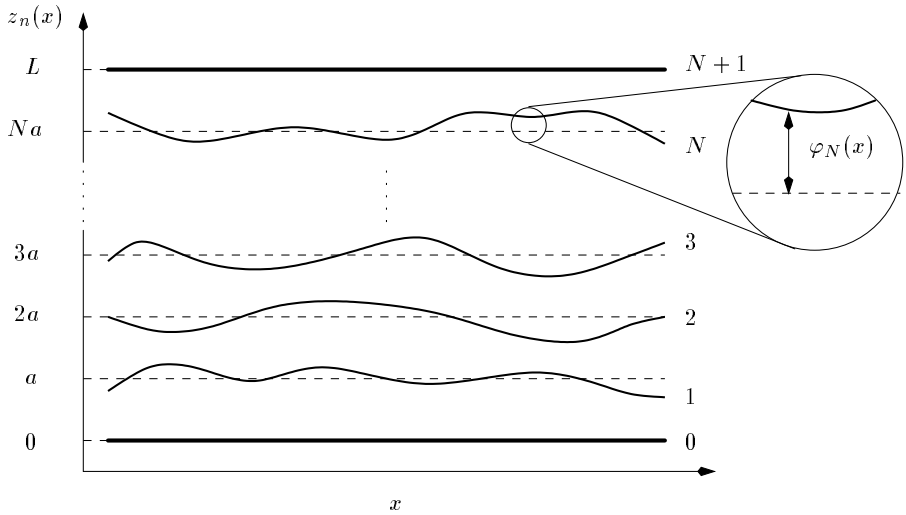}}\vspace{5mm}
\caption{\label{sst} Stack of $N$ strings with equilibrium spacing $a$ between two walls of 
distance $L$. The magnifier shows the local displacement field $\varphi_N(x)$ as
the distance from the position $Na$. 
The walls are labeled by $0$ and $N+1$ and treated as nonfluctuating strings with
$\varphi_0(x)\equiv\varphi_{N+1}(x)\equiv 0$.}
\end{figure}

We begin by studying the exactly solvable statistical properties of a
stack of $N$ almost parallel strings in a plane, which are not allowed to cross each other
and whose average spacing at low temperature is $a$.
The system is enclosed between parallel line-like walls 
with a separation $L$ as illustrated in Fig.~\ref{sst}. In the Monge parametrization,
the vertical position of a point of the $n$th string is $z_n=z_n(x)$. 
Since the vertical positions of the $n$th string are
fluctuating around the low-temperature equilibrium position at $na$, 
it is useful to introduce the displacement
fields $\varphi_n(x)$
\begin{equation}
\label{eq:pos}
\varphi_n(x)\equiv z_n(x)-n a.
\end{equation} 
The thermodynamic partition function is given by the  
functional integral
\begin{equation}
\label{eq:part}
Z^{\rm s}=\prod\limits_{n=1}^N\prod\limits_x\left[
\int\limits_{\varphi_{n-1}(x)+a}^{\varphi_{n+1}(x)-a} 
\frac{d\varphi_n(x)}{\sqrt{2\pi k_BT/\sigma}}\right]\,\exp\left\{-\frac{\sigma}{2k_BT} 
\sum\limits_{n=1}^N\int_{-\infty}^\infty dx\,\left[\frac{d\varphi_n(x)}{dx} \right]^2
\right\},
\end{equation}
where $\sigma$ is the string tension, $T$ is the temperature, and $k_B$ the Boltzmann factor.
We are interested in the free energy per unit length 
\begin{equation}
\label{eq:freex}
f^{\rm s}_N\equiv -\frac{k_BT}{A}\,{\rm ln}\,Z^{\rm s},
\end{equation}
with $A=\int_{-\infty}^\infty dx$.
Since the strings may not pass through each other, the fluctuations $\varphi_n(x)$
of the $n$th string are restricted to the interval
\begin{equation}
\label{restr} 
\varphi_n(x)\in \{\varphi_{n-1}(x)+a,\varphi_{n+1}(x)-a\}.
\end{equation} 
\subsection{Free Fermion Model}
The restriction (\ref{restr})  makes it 
difficult to solve the functional integral (\ref{eq:part}) explicitly. 
It is, however, possible to find a solution using an alternative of the same systems as
a $1+1$-dimensional 
Fermi gas observed by de Gennes~\cite{deGennes}. Using this analogy,
Gompper and Kroll~\cite{GK} determined the
$1/a^2$-contribution to $\Delta f^{\rm s}_N$ 
relevant for the pressure law~(\ref{press}) as
\begin{equation}
\label{eq:freexn}
\Delta f^{\rm s}_N=\alpha^{\rm s}_N \frac{(k_BT)^2}{\sigma a^2},
\end{equation}
with the pressure constants
\begin{equation}
\label{eq:consts}
\alpha^{\rm s}_N=\frac{\pi^2}{12}\frac{2N+1}{N+1}.
\end{equation}
For $N\to\infty$, this constant has the finite limit $\alpha_\infty^s=\pi^2/6$.
The analogy with fermions cannot be used
to calculate the free energy of a stack 
of membranes, where only approximate methods are available. We shall use a strong-coupling
theory as in Ref.~\cite{mem}. As a preparation, we we  
apply this theory to the exactly solvable system of a stack of strings.
\subsection{Perturbative Approach}
The difficulty in solving 
the functional integral (\ref{eq:part}) arises from the restriction (\ref{restr})
of the fluctuations
by the neighbouring strings. To deal with this strong repulsion, we
introduce into the action of the functional integral (\ref{eq:part})
a smooth potential which keeps the 
strings apart in such a way that the integration
interval for the fluctuations can be extended to $\varphi_n(x)\in\{-\infty,\infty\}$.
At the end we take a strong-coupling limit which ensures (\ref{restr}).
In Ref.~\cite{mem}, such a method was used to evaluate the pressure constant for
one membrane between walls. The smooth potential
for the analogous case of one string is
$V(\varphi(x))=(2a\,\mu/\pi)^2\tan^2[\pi\varphi(x)/2a]$ which describes the hard walls
{\it exactly} for $\mu\to 0$. This potential is symmetric and possesses a minimum 
at $\varphi(x)=0$.
Thus its Taylor expansion around this minimum is a series in even powers of $\varphi(x)$.

In the case of $N$ strings, the minima of the repulsion potential 
should lie at the equlibrium positions of the strings. 
The Taylor expansion of such a potential will also 
have terms with odd powers. Unlike the one-string system, where fluctuations
are limited by fixed walls, the range of the displacements $\varphi_n(x)$ 
of the $n$th string in an 
$N$-string system depends on the positions $z_{n-1}(x)$ and $z_{n+1}(x)$
of the neighbouring strings. Thus the
potential will be taken as a sum
\begin{equation}
\label{eq:pot}
V_{\rm eff}(z_0(x),z_1(x),\ldots,z_N(x),z_{N+1}(x))=\frac{\sigma}{2}
\sum\limits_{n=1}^{N+1}V_\mu(\onabla_n z_n(x)),
\end{equation}
where $\onabla_n z_n(x)$ denotes the pre-point lattice gradient $z_n(x)-z_{n-1}(x)$.
This potential includes the interaction of the first and last 
strings with the walls as
non-fluctuating strings at $z_0=0$ and $z_{N+1}=(N+1)a=L$: 
\begin{equation}
\label{eq:bound}
\varphi_0(x)=\varphi_{N+1}(x)=0.  
\end{equation}
In the limit $\mu\to 0$, the potential $V_\mu(\onabla_n z_n(x))$ should again yield
an infinitely strong repulsion of two neighbouring strings for $z_n(x)$ close to $z_{n-1}(x)$. 
For $z_n(x)>z_{n-1}(x)$, the limiting potential should be zero.
\begin{figure}
\centerline{\epsfxsize=12cm \epsfbox{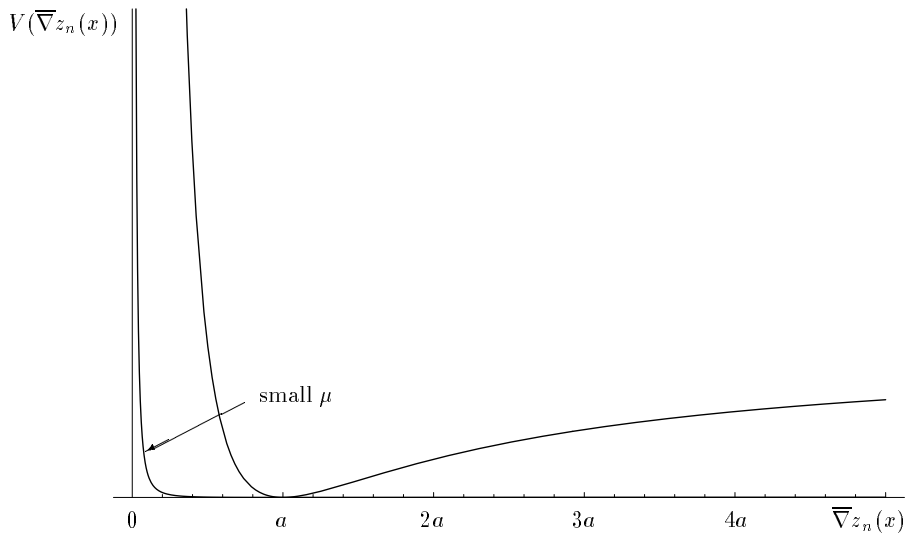}}\vspace{5mm}
\caption{\label{pots} Potential $V_\mu(\onabla_n z_n(x))$ of string-string interaction
for finite $\mu$ and small $\mu$ as function of $\onabla_n z_n(x)=z_n(x)-z_{n-1}(x)$.
The strings repel each other strongly for $\onabla_n z_n(x)\to 0$, while
the potential has a minimuim at the equilibrium separation $\onabla_n z_n(x)=a$, and we 
choose to normalize it to zero at that point. 
}
\end{figure}
As a matter of choice, we let
the potential between two strings $V_\mu(\onabla_n z_n(x))$ be minimal
and zero
at the positions $z_n^{\rm eq}=na$ and
$z_{n-1}^{\rm eq}=a(n-1)$: $dV_\mu(a)/d\, \onabla_n z_n(x)=0$ and 
$V_\mu(z_n^{\rm eq}-z_{n-1}^{\rm eq})=V_\mu(a)=0$ (see Fig.~\ref{pots}). 

The Taylor expansion around the miminum is, in terms of the variables (\ref{eq:pos}),
\begin{equation}
\label{eq:Taylor}
V_\mu(\onabla_n\varphi_n(x))=\frac{\mu^2}{2}[\onabla_n\varphi(x)]^2+\mu^2\sum\limits_{k=1}^{N+1}c_k g^k
[\onabla_n\varphi(x)]^{k+2}.
\end{equation}
The parameter $\mu$ governs the harmonic term.
The introduction of the coupling constant $g=1/a$
makes the coefficients $c_k$ dimensionless, and
the partition function (\ref{eq:part}) becomes
\begin{eqnarray}
  \label{eq:part2}
  Z^{\rm s}=\lim_{\mu\to 0}&&\oint {\cal D}^N\varphi(x)\,\exp\left\{-\frac{\sigma}{2k_BT}
\sum\limits_{n=1}^{N+1}\int\limits_{-\infty}^\infty dx\,\left(\left[\frac{d\varphi_n(x)}{dx} 
\right]^2+\frac{1}{2}\mu^2\left[\onabla_n \varphi_n(x) \right]^2 
\right\}\right)\nonumber\\
&&\times\exp\left\{-\frac{\sigma}{2k_BT}\mu^2\sum\limits_{k=1}^\infty c_kg^k
\sum\limits_{n=1}^{N+1}\int\limits_{-\infty}^\infty dx
\,[\onabla_n\varphi_n(x)]^{k+2}\right\}
\end{eqnarray}
with the integral measure
\begin{equation}
  \label{eq:meas}
  \oint {\cal D}^N\varphi(x)=\prod\limits_{n=1}^N\prod\limits_x
\left[\int_{-\infty}^{\infty} 
\frac{d\varphi_n(x)}{\sqrt{2\pi k_BT/\sigma}} \right].
\end{equation}
The harmonic part of the partition function can be written as
\begin{equation}
\label{eq:pex}
Z^{\rm s}_\mu=\oint {\cal D}^N\varphi(x)\,\exp\left\{-\frac{1}{2}
\sum\limits_{n=1}^{N+1}\sum\limits_{n'=1}^{N+1}\int\limits_{-\infty}^\infty dx
\int\limits_{-\infty}^\infty dx'\,\varphi_n(x)[G^{\rm s}_{nn'}(x,x')]^{-1}\varphi_{n'}(x')\right\}
\end{equation}
with the functional matrix
\begin{equation}
\label{eq:kern}
[G^{\rm s}_{nn'}(x,x')]^{-1}=-\frac{\sigma}{k_BT}\left(\frac{d^2}{dx^2}+
\frac{1}{2}\mu^2\,\onabla_n\nabla_n
\right)\delta(x-x')\delta_{nn'}.
\end{equation}
Here $\nabla_n\varphi_n(x)=\varphi_{n+1}(x)-\varphi_n(x)$ denotes the post-point lattice gradient
in the $z$-direction, and $\onabla_n\nabla_n$ is the lattice version of the Laplace 
operator~\cite{PI}. 

Let us now impose the vanishing
of the fluctuations of the walls (\ref{eq:bound}), corresponding
to Dirichlet boundary conditions. For
a finite number $N$ of strings, the Kronecker symbol $\delta_{n\,n'}$ in Eq.~(\ref{eq:kern})
has the Fourier representation
\begin{equation}
\label{eq:kdelta}
\delta_{nn'}=\frac{2}{N+1}\sum\limits_{m=1}^N\sin\nu_mna\,\sin\nu_mn'a
\end{equation}
with wave numbers $\nu_m=\pi m/(N+1)a$. Thus the kernel $[G^{\rm s}_{n\,n'}(x,x')]^{-1}$ 
may be written
in Fourier space as
\begin{equation}
\label{eq:kfou}
[G^{\rm s}_{nn'}(x,x')]^{-1}=\frac{2}{N+1}\sum\limits_{m=1}^N\sin\nu_mna\,\sin\nu_mn'a
\int\limits_{-\infty}^\infty\frac{dk}{2\pi}[G^{\rm s}_m(k)]^{-1}e^{-ik(x-x')}
\end{equation}
with the Fourier components
\begin{equation}
\label{eq:gfou}
[G^{\rm s}_m(k)]^{-1}=\frac{\sigma}{k_BT}\left[k^2+2\mu^2\sin^2(\nu_ma/2)\right].
\end{equation}
Integrating over $k$ in the spectral representation 
(\ref{eq:kfou}) leads immediately to the correlation
function in configuration space
\begin{equation}
\label{eq:g}
G^{\rm s}_{nn'}(x,x')=\frac{1}{\sqrt{2}(N+1)}\frac{k_BT}{\mu\sigma}\sum\limits_{m=1}^N
\frac{\sin\nu_mna\sin\nu_mn'a}{\sin(\nu_ma/2)}\,e^{-\sqrt{2}\mu|x-x'|\sin(\nu_ma/2)},
\end{equation}
and to the harmonic partition function (\ref{eq:pex})
\begin{equation}
\label{pex2}
Z^{\rm s}_\mu=\exp\left\{-\frac{1}{2}{\rm Tr}\,{\rm ln}\,[G^{\rm s}]^{-1} \right\}=
e^{-Af^{\rm s}_{N,\mu}/k_BT},
\end{equation}
the exponent giving the free energy per length:
\begin{equation}
\label{fex}
f^{\rm s}_{N,\mu}=\mu\,\frac{k_BT}{2}\frac{\sin[\pi N/4(N+1)]}{\sin[\pi/4(N+1)]},
\end{equation}
which vanishes for $\mu= 0$.

The full partition function $Z^{\rm s}$
in Eq.~(\ref{eq:part2}) is now calculated perturbatively. We introduce harmonic 
expectation values
\begin{equation}
\label{eq:exp}
\means{~\cdots~}=\left[Z^{\rm s}_\mu \right]^{-1}\oint {\cal D}^N\varphi(x)\,\cdots\,
\exp\left\{ -\frac{1}{2}
\sum\limits_{n=1}^{N+1}\sum\limits_{n'=1}^{N+1}\int\limits_{-\infty}^\infty dx
\int\limits_{-\infty}^\infty dx'\,\varphi_n(x)[G^{\rm s}_{nn'}(x,x')]^{-1}\varphi_{n'}(x')\right\}
\end{equation}
in terms of which the correlation function is given by
\begin{equation}
\label{eq:corr}
G^{\rm s}_{nn'}(x,x')=\means{\varphi_n(x)\varphi_{n'}(x')}.
\end{equation}
The perturbation expansion contains
the two-point correlation function of $\onabla_n\varphi_n(x)$ which is given by
\begin{equation}
\label{eq:ncorr}
\means{\onabla_n\varphi_n(x)\onabla_{n'}\varphi_{n'}(x')}=\onabla_n\onabla_{n'}\,
G^{\rm s}_{nn'}(x,x').
\end{equation}
We now expand the second exponential
in Eq.~(\ref{eq:part2}) in powers of the coupling constant $g$. 
Harmonic expectation values
with odd powers of $\onabla_n\varphi_n(x)$ do not contribute, and the expansion reads:
\begin{eqnarray}
\label{eq:partex}
Z^{\rm s}=\lim_{\mu\to 0}Z^{\rm s}_\mu&&\left[1-g^2\left(\frac{\sigma}{2k_BT}\mu^2c_2
\sum\limits_{n=1}^{N+1}\int\limits_{-\infty}^\infty dx\means{[\onabla_n\varphi_n(x)]^4}
\right.\right.\nonumber\\
&&-\left.\left.\frac{1}{2!}\frac{\sigma^2}{4k_B^2T^2}\mu^4c_1^2\sum\limits_{n,n'=1}^{N+1}
\int\limits_{-\infty}^\infty dx\int\limits_{-\infty}^\infty dx'\,
\means{[\onabla_n\varphi_n(x)]^3[\onabla_{n'}\varphi_{n'}(x')]^3} \right)+\ldots \right].
\end{eqnarray} 
In the sequel, we restrict ourselves to the terms of second order in $g=1/a$, which 
contribute directly to the pressure law as in Eq.~(\ref{eq:freexn}).
The higher powers diverge for $\mu\to 0$, and in Ref.~\cite{mem} it was shown how to calculate
from them a finite strong-coupling limit. Here we shall ignore these terms for reasons to
be explained shortly.
Reexpressing the right-hand side of Eq.~(\ref{eq:partex}) as an exponential of 
a cumulant expansion,
we obtain the free energy per length
\begin{eqnarray}
\label{eq:freep}
f^{\rm s}_N&=&\lim_{\mu\to 0}g^2\left(\frac{\sigma}{2k_BT}\mu^2c_2
\sum\limits_{n=1}^{N+1}\int\limits_{-\infty}^\infty dx\cums{[\onabla_n\varphi_n(x)]^4}
\right.\nonumber\\
&&-\left.\frac{1}{2!}\frac{\sigma^2}{4k_B^2T^2}\mu^4c_1^2\sum\limits_{n,n'=1}^{N+1}
\int\limits_{-\infty}^\infty dx\int\limits_{-\infty}^\infty dx'\,
\cums{[\onabla_n\varphi_n(x)]^3[\onabla_{n'}\varphi_{n'}(x')]^3} \right)+\ldots~.
\end{eqnarray}
We have used that the free energy $f^{\rm s}_{N,\mu}$ of the harmonic system (\ref{fex})
vanishes in the limit $\mu\to 0$. The first cumulants are
the expectations
\begin{eqnarray}
\label{cums}
\cums{O_1(\onabla\varphi_{n_1}(x_1))}&=&\means{O_1(\onabla\varphi_{n_1}(x_1))}\nonumber \\
\cums{O_1(\onabla\varphi_{n_1}(x_1))O_2(\onabla\varphi_{n_2}(x_2))}
&=&\means{O_1(\onabla\varphi_{n_1}(x_1))O_2(\onabla\varphi_{n_2}(x_2))}-
\means{O_1(\onabla\varphi_{n_1}(x_1))}\means{O_2(\onabla\varphi_{n_2}(x_2))},\nonumber \\
&\vdots&\quad,
\end{eqnarray}
defined for any polynomial function $O_i(\onabla\varphi_{n_i}(x_i))$  
of $\onabla\varphi_{n_i}(x_i)$.
Following Wick's rule, we expand the expectations 
into products of two-point correlation functions (\ref{eq:ncorr}). 
The different terms are displayed with the help of Feynman diagrams, in which
lines and vertices represent the correlation functions and interactions:
\begin{fmffile}{bkpdiag}
\begin{eqnarray}
\parbox{30mm}{\centerline{
\begin{fmfgraph*}(20,8)
\setval
\fmfleft{v1}
\fmfright{v2}
\fmf{plain}{v1,v2}
\fmflabel{$x_1,n_1$}{v1}
\fmflabel{$x_2,n_2$}{v2}
\end{fmfgraph*}
}} ~~&\longrightarrow &~~ 
\means{\onabla_{n_1}\varphi_{n_1}(x_1)\onabla_{n_2}\varphi_{n_2}(x_2)},\\
\parbox{30mm}{\centerline{
\begin{fmfgraph}(8,8)
\setval
\fmfforce{0.5w,0.5h}{v1}
\fmfdot{v1}
\end{fmfgraph}
}}~~&\longrightarrow &~~ \sum\limits_{n=1}^{N+1}\int_{-\infty}^{\infty}dx.
\end{eqnarray}
In what follows, we assume that the potential parameters $c_k$ with $k\ge 3$ are chosen in 
such a way that they make all terms of order $g^3$ and higher vanish. Dividing the
free energy (\ref{eq:freep}) by $N$, we obtain the following expression for the free energy
per length and string which can be compared with Eq.~(\ref{eq:freexn}):
\begin{eqnarray}
\label{freeexp}
\Delta f^{\rm s}_N=\lim_{\mu\to 0}\,\left\{
\frac{3}{2}\frac{\sigma\mu^2}{A a^2}c_2\;
\parbox{13mm}{\centerline{
\begin{fmfgraph}(10mm,5mm)
\setval
\fmfforce{0.5w,0.5h}{v}
\fmfi{plain}{fullcircle scaled 1h shifted (1/4w,1/2h)}
\fmfi{plain}{fullcircle scaled 1h shifted (3/4w,1/2h)}
\fmfdot{v}
\end{fmfgraph}
}}-\frac{1}{8}\frac{\sigma^2 \mu^4}{k_BT A a^2}c_1^2\left(6\;
\parbox{9mm}{\centerline{
\begin{fmfgraph}(6mm,6mm)
\setval
\fmfforce{0w,0.5h}{v1}
\fmfforce{1w,0.5h}{v2}
\fmf{plain,left=1}{v1,v2,v1}
\fmf{plain}{v1,v2}
\fmfdot{v1,v2}
\end{fmfgraph}
}}+9\;
\parbox{18mm}{\centerline{
\begin{fmfgraph}(15mm,7mm)
\setval
\fmfforce{0.33w,0.5h}{v1}
\fmfforce{0.66w,0.5h}{v2}
\fmf{plain}{v1,v2}
\fmfi{plain}{reverse fullcircle scaled 0.33w shifted (0.165w,0.5h)}
\fmfi{plain}{fullcircle rotated 180 scaled 0.33w shifted (0.825w,0.5h)}
\fmfdot{v1,v2}
\end{fmfgraph}
}}
\right)\right\}.
\end{eqnarray}
The calculation of the Feynman diagrams is straightforward 
using Eq.~(\ref{eq:g}). The evaluation is only complicated by the Dirichlet
boundary conditions, which destroy momentum conservation. This makes
the numeric calculation quite time-consuming for an
increasing number $N$ of strings. As an explicit example consider the
sunset diagram which requires the evaluation of the multiple sum
\begin{eqnarray}
\label{eq:sun}
\parbox{9mm}{\centerline{
\begin{fmfgraph}(6mm,6mm)
\setval
\fmfforce{0w,0.5h}{v1}
\fmfforce{1w,0.5h}{v2}
\fmf{plain,left=1}{v1,v2,v1}
\fmf{plain}{v1,v2}
\fmfdot{v1,v2}
\end{fmfgraph}
}}&\equiv& A\frac{k_B^3T^3}{\mu^4}\frac{1}{2(N+1)^3}\sum\limits_{n_1,n_2=1}^{N+1}
\sum\limits_{m1,m2,\atop m3=1}^{N} h_{n_1n_2}^{m_1}h_{n_1n_2}^{m_2}h_{n_1n_2}^{m_3}\nonumber\\
&&\times \frac{1}
{\sin(\nu_{m_1}a/2)\,\sin(\nu_{m_2}a/2)\,\sin(\nu_{m_3}a/2)\,[\sin(\nu_{m_1}a/2)+
\sin(\nu_{m_2}a/2)+\sin(\nu_{m_3}a/2)]}
\end{eqnarray}
with the abbreviation
\begin{eqnarray}
\label{abbrev}
h_{n_1n_2}^{m}&=&\sin \nu_mn_1 a\,\sin\nu_m n_2 a-\sin \nu_mn_1 a\,\sin\nu_m (n_2-1) a
\nonumber\\
&&-\sin \nu_m(n_1-1) a\,\sin\nu_m n_2 a+\sin \nu_m(n_1-1) a\,\sin\nu_m (n_2-1) a.
\end{eqnarray}
It is useful to factor out the physical dimension of the diagram. Any Feynman integral
$W^{\rm s}$ with $l$ lines and $v$ vertices can be expressed in terms of a reduced dimensionless
Feynman integral $W^{{\rm s},r}$ as 
\begin{equation}
\label{dim}
W^{\rm s}=A\left(\frac{k_BT}{\sigma}\right)^l \mu^{-(l+v-1)}\, W^{{\rm s},r}.
\end{equation}
This brings Eq.~(\ref{freeexp}) to the form
\begin{eqnarray}
\label{eq:freeexp}
\Delta f^{\rm s}_{N}&=&\alpha^{\rm s}_{N}\frac{k_B^2T^2}{\sigma a^2},\\
\label{consts}
\alpha^{\rm s}_{N}&=&\frac{1}{N}\left[\frac{3}{2}c_2 \;
\parbox{13mm}{\centerline{
\begin{fmfgraph}(10mm,5mm)
\setval
\fmfforce{0.5w,0.5h}{v}
\fmfforce{0.5w,1h}{vl}
\fmfi{plain}{fullcircle scaled 1h shifted (1/4w,1/2h)}
\fmfi{plain}{fullcircle scaled 1h shifted (3/4w,1/2h)}
\fmfdot{v}
\end{fmfgraph}
}}^{{\rm s},r}
-c_1^2\left(\frac{3}{4}\;
\parbox{9mm}{\centerline{
\begin{fmfgraph}(6mm,6mm)
\setval
\fmfforce{0w,0.5h}{v1}
\fmfforce{1w,0.5h}{v2}
\fmf{plain,left=1}{v1,v2,v1}
\fmf{plain}{v1,v2}
\fmfdot{v1,v2}
\end{fmfgraph}
}}^{{\rm s},r}
+\frac{9}{8}\;
\parbox{18mm}{\centerline{
\begin{fmfgraph}(15mm,7mm)
\setval
\fmfforce{0.33w,0.5h}{v1}
\fmfforce{0.66w,0.5h}{v2}
\fmf{plain}{v1,v2}
\fmfi{plain}{reverse fullcircle scaled 0.33w shifted (0.165w,0.5h)}
\fmfi{plain}{fullcircle rotated 180 scaled 0.33w shifted (0.825w,0.5h)}
\fmfdot{v1,v2}
\end{fmfgraph}
}}^{{\rm s},r}
\right)\right],
\end{eqnarray}
where the diagrams indicate the reduced
Feynman integrals. Their values are listed
in Table~\ref{mtab1} 
for different string numbers $N$. Note that the $1/a^2$-contributions to the
free energy per length and string in Eq.~(\ref{freeexp}) is independent
of $\mu$ since the $\mu$-prefactors are cancelled by the $\mu$-dependence of the diagrams.
Thus the limit $\mu\to 0$ becomes trivial for these contributions.

With the knowledge of the exact values of the constants 
$\alpha^{\rm s}_N$ from Eq.~(\ref{eq:consts}), 
we are now in a position to determine the potential parameters $c_1$ and $c_2$ from 
Eq.~(\ref{consts}) to obtain the exact result from the two-loop expansion Eq.~(\ref{consts}). 
Comparing Eqs.~(\ref{consts}) and (\ref{eq:consts}) for $N=1$ and $N=2$, we obtain
\begin{equation}
\label{potparam}
c_1=\frac{\pi}{3},\quad c_2=\frac{\pi^2}{6}.
\end{equation}
Note that (\ref{consts}) consists of more equations than necessary to compute $c_1$ and $c_2$.
It turns out, however, that all of them give the same $c_1$ and $c_2$, such that the same 
potential (\ref{eq:Taylor}) can be used for any $N$.
This is the essential 
basis for applying this procedure to a stack of membranes.

We now justify the neglect of the higher $g$ powers which would in principle give a further 
contribution to the pressure constant $\alpha_N^{\rm s}$ in the strong-coupling limit. We
simply observe that it is possible to choose the higher expansion coefficients $c_k$
to make all higher $g^n$ contributions vanish~\cite{kc}.
\section{Stack of Membranes}
\begin{figure}
\centerline{\epsfxsize=9cm \epsfbox{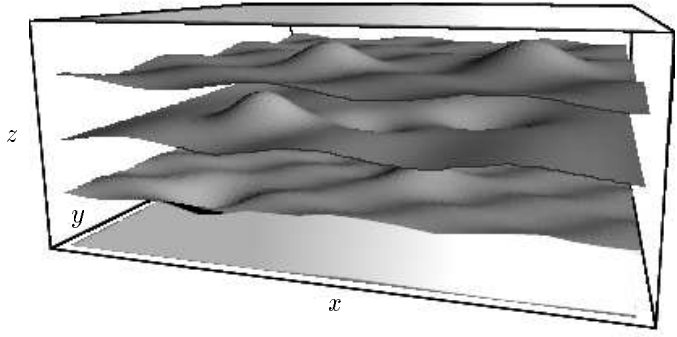}}
\caption{\label{mst} Stack of selfavoiding fluid membranes fluctuating 
in $z$-direction between two walls. 
As for the previous stack of strings, the walls are treated as nonfluctuating membranes.}
\end{figure}

Having determined 
the parameters $c_1$ and $c_2$ of the Taylor expansion (\ref{eq:Taylor})
of the smooth potential applicable for any 
number of strings, we shall now use the same potential for a perturbative expansion in a
stack of $N$ membranes displayed in Fig.~\ref{mst}.
The equilibrium spacing at low temperature between the
membranes is again $a$. Denoting the vectors in the plane by 
${\bf x}=(x,y)$, the vertical displacements of the membranes from the positions $na$ are
$\varphi_n({\bf x})$, with Dirichlet boundary conditions at $z_0$ and $z_{N+1}$
\begin{equation}
\label{eq:mdbc}
\varphi_0({\bf x})=\varphi_{N+1}({\bf x})=0.
\end{equation} 
For membranes without tension, the energy has the harmonic approximation
\begin{equation}
\label{eq:mcurv}
E_{C,n}=\frac{\kappa}{2}\int d^2x\,\left[\partial^2 \varphi_n({\bf x})\right]^2,
\end{equation}
where $\kappa$ is the bending stiffness and $\partial^2=\partial_x^2+\partial_y^2$ 
is the Laplacian
in the plane parallel to the walls.
By analogy with the previous section, 
the kernel of the harmonic stack reads now
\begin{equation}
\label{eq:mkern}
[G_{n_1n_2}({\bf x}_1,{\bf x}_2)]^{-1}=-\frac{\kappa}{k_BT}
\left(\left[\partial_{{\bf x}_1}^2\right]^2+\frac{1}{2}\mu^4\onabla_{n_1}\nabla_{n_1}
\right)\delta({\bf x}_1-{\bf x}_2)\delta_{n_1n_2},
\end{equation}
where we have used a mass parameter $\mu^4$ instead of $\mu^2$, for dimensional reasons.
The partition function for the stack of membranes is then written up to order $g^2=1/a^2$ by
\begin{eqnarray}
\label{eq:pmem}
Z=\lim_{\mu\to 0}&&\oint {\cal D}^N\varphi({\bf x})\,\exp\left\{-\frac{\kappa}{2k_BT} 
\sum\limits_{n_1,n_2=1}^{N+1}\int d^2x_1\int d^2x_2\,\varphi_{n_1}({\bf x}_1)
[G_{n_1n_2}({\bf x}_1,{\bf x}_2)]^{-1}\varphi_{n_2}({\bf x}_2)\right\}\nonumber\\
&&\hspace{-1cm}\times\left[1-g^2\left(\frac{\kappa}{2k_BT}\mu^4c_2\sum\limits_{n=1}^{N+1}\int d^2x
\left[\onabla\varphi_n({\bf x}) \right]^4-\frac{\kappa^2}{8k_B^2T^2}\mu^8c_1^2
\sum\limits_{n_1,n_2=1}^{N+1}\int d^2x_1\int d^2x_2\,
\left[\onabla\varphi_{n_1}({\bf x}_1)\right]^3\left[ \onabla\varphi_{n_2}({\bf x}_2) \right]^3
\right) \right],\nonumber\\
&&\quad
\end{eqnarray}
with the same parameters $c_1$ and $c_2$ of the Taylor expansion (\ref{eq:Taylor})
as in the string system, determined in
Eq.~(\ref{potparam}). 
We neglect terms of order $g^3$ which certainly contribute in the strong-coupling limit,
and which vanish only for the strings, where the partition function 
(\ref{eq:partex}) with the choice (\ref{potparam}) for the parameters $c_1$, $c_2$
is exact in second order. 
An evaluation of the neglected terms by variational perturbation theory is expected to
give only a negligible contribution to our final result. 

Inverting the kernel (\ref{eq:mkern}) yields
the correlation function
\begin{equation}
\label{eq:mg}
G_{n_1n_2}({\bf x}_1,{\bf x}_2)=\frac{2}{N+1}\frac{k_BT}{\kappa}\sum\limits_{m=1}^N
\sin\nu_mn_1a\sin\nu_mn_2a
\int\frac{d^2k}{(2\pi)^2}\,\frac{1}{k^4+2\mu^4\sin^2(\nu_ma/2)}e^{-i{\bf k}({\bf x}_1-{\bf x}_2)}.
\end{equation}
The explicit calculation of the Fourier integral leads to a difference of modified
Bessel functions $K_0(x)$ as in Ref.~\cite{mem}:
\begin{eqnarray}
\label{eq:mg2}
G_{n_1n_2}({\bf x}_1,{\bf x}_2)&=&\frac{i}{\sqrt{8}\pi(N+1)\mu^2}\frac{k_BT}{\kappa}
\sum\limits_{m=1}^N
\frac{\sin\nu_mn_1a\sin\nu_mn_2a}{\sin(\nu_ma/2)}\nonumber\\
&&\times\left[K_0\left(2^{1/4}\sqrt{i\sin(\nu_ma/2)}\mu|{\bf x}_1-{\bf x}_2|\right)-
K_0\left(2^{1/4}\sqrt{-i\sin(\nu_ma/2)}\mu|{\bf x}_1-{\bf x}_2|\right) \right].
\end{eqnarray}
For ${\bf x}_1={\bf x}_2\equiv {\bf x}$ and $n_1=n_2\equiv n$, this reduces to
\begin{equation}
\label{eq:mgx}
G_{nn}({\bf x},{\bf x})=\frac{1}{\sqrt{32}(N+1)\mu^2}\frac{k_BT}{\kappa}\sum\limits_{m=1}^N
\frac{\sin^2\nu_mna}{\sin(\nu_ma/2)},
\end{equation}
leading to the partition function of the harmonic system
\begin{equation}
\label{eq:pexm}
Z_\mu=\exp\left\{-\frac{1}{2}{\rm Tr}\,{\rm ln}\,G^{-1}\right\}=
\exp\left\{-\mu^2\frac{A}{8}\frac{\sin[\pi N/4(N+1)]}{\sin[\pi/4(N+1)]} \right\},
\end{equation}
where $A=\int d^2x$ is the area of the projected plane of the membranes.
The free energy per area $f_{N,\mu}=-(k_BT/A){\rm ln}\,Z_\mu$ vanishes
again for $\mu =0$.

As for the calculation of the free energy of the stack of strings, we introduce harmonic
expectation values 
\begin{equation}
\label{eq:expm}
\meanm{~\cdots~}=\left[Z_\mu \right]^{-1}\oint {\cal D}^N\varphi({\bf x})\,\cdots\,
\exp\left\{ -\frac{1}{2}
\sum\limits_{n=1}^{N+1}\sum\limits_{n'=1}^{N+1}\int\limits_{-\infty}^\infty d^2x
\int\limits_{-\infty}^\infty d^2x'\,
\varphi_n({\bf x})[G_{nn'}({\bf x},{\bf x}')]^{-1}\varphi_{n'}({\bf x}')\right\}
\end{equation}
which appear in the perturbation expansion of
(\ref{eq:pmem}), the cumulants yielding 
a perturbative expansion for the free energy per area $f_N=-(k_BT/A){\rm ln}\,Z$.
The lines and vertices in the 
Feynman diagrams stand now for
\begin{eqnarray}
\parbox{30mm}{\centerline{
\begin{fmfgraph*}(20,8)
\setval
\fmfleft{v1}
\fmfright{v2}
\fmf{plain}{v1,v2}
\fmflabel{${\bf x}_1,n_1$}{v1}
\fmflabel{${\bf x}_2,n_2$}{v2}
\end{fmfgraph*}
}}~~ &\longrightarrow &~~ 
\meanm{\onabla_{n_1}\varphi_{n_1}({\bf x}_1)\onabla_{n_2}\varphi_{n_2}({\bf x}_2)}\\
\parbox{30mm}{\centerline{
\begin{fmfgraph}(8,8)
\setval
\fmfforce{0.5w,0.5h}{v1}
\fmfdot{v1}
\end{fmfgraph}
}}~~ &\longrightarrow &~~ \sum\limits_{n=1}^{N+1}\int d^2x,
\end{eqnarray}
and the two-loop approximation to the free energy per area and membrane in order $1/a^2$ reads
\begin{eqnarray}
\label{freeexpm}
\Delta f_N=\lim_{\mu\to 0}\,\left\{
\frac{\pi^2}{4}\frac{\kappa\mu^4}{A a^2}\;
\parbox{13mm}{\centerline{
\begin{fmfgraph}(10mm,5mm)
\setval
\fmfforce{0.5w,0.5h}{v}
\fmfi{plain}{fullcircle scaled 1h shifted (1/4w,1/2h)}
\fmfi{plain}{fullcircle scaled 1h shifted (3/4w,1/2h)}
\fmfdot{v}
\end{fmfgraph}
}}-\frac{\pi^2\kappa^2 \mu^8}{k_BT A a^2}\left(\frac{1}{12}\;
\parbox{9mm}{\centerline{
\begin{fmfgraph}(6mm,6mm)
\setval
\fmfforce{0w,0.5h}{v1}
\fmfforce{1w,0.5h}{v2}
\fmf{plain,left=1}{v1,v2,v1}
\fmf{plain}{v1,v2}
\fmfdot{v1,v2}
\end{fmfgraph}
}}+\frac{1}{8}\;
\parbox{18mm}{\centerline{
\begin{fmfgraph}(15mm,7mm)
\setval
\fmfforce{0.33w,0.5h}{v1}
\fmfforce{0.66w,0.5h}{v2}
\fmf{plain}{v1,v2}
\fmfi{plain}{reverse fullcircle scaled 0.33w shifted (0.165w,0.5h)}
\fmfi{plain}{fullcircle rotated 180 scaled 0.33w shifted (0.825w,0.5h)}
\fmfdot{v1,v2}
\end{fmfgraph}
}}
\right)\right\}.
\end{eqnarray}
Going over to reduced Feynman integrals as in Eq.~(\ref{dim})
\begin{equation}
\label{dimm}
W=A\left(\frac{k_BT}{\kappa}\right)^l \mu^{-2(l+v-1)}\, W^{r},
\end{equation}
where $v$ is the number of vertices and $l$ the number of lines of the diagram,
we obtain
\begin{eqnarray}
\label{eq:freeexpm}
\Delta f_{N}&=&\alpha_{N}\frac{k_B^2T^2}{\kappa a^2},\\
\label{constsm}
\alpha_{N}&=&\frac{\pi^2}{N}\left(\frac{1}{4} \;
\parbox{13mm}{\centerline{
\begin{fmfgraph}(10mm,5mm)
\setval
\fmfforce{0.5w,0.5h}{v}
\fmfforce{0.5w,1h}{vl}
\fmfi{plain}{fullcircle scaled 1h shifted (1/4w,1/2h)}
\fmfi{plain}{fullcircle scaled 1h shifted (3/4w,1/2h)}
\fmfdot{v}
\end{fmfgraph}
}}^{r}
-\frac{1}{12}\;
\parbox{9mm}{\centerline{
\begin{fmfgraph}(6mm,6mm)
\setval
\fmfforce{0w,0.5h}{v1}
\fmfforce{1w,0.5h}{v2}
\fmf{plain,left=1}{v1,v2,v1}
\fmf{plain}{v1,v2}
\fmfdot{v1,v2}
\end{fmfgraph}
}}^{r}
-\frac{1}{8}\;
\parbox{18mm}{\centerline{
\begin{fmfgraph}(15mm,7mm)
\setval
\fmfforce{0.33w,0.5h}{v1}
\fmfforce{0.66w,0.5h}{v2}
\fmf{plain}{v1,v2}
\fmfi{plain}{reverse fullcircle scaled 0.33w shifted (0.165w,0.5h)}
\fmfi{plain}{fullcircle rotated 180 scaled 0.33w shifted (0.825w,0.5h)}
\fmfdot{v1,v2}
\end{fmfgraph}
}}^{r}
\right).
\end{eqnarray}
The pressure exerted by the membranes upon the walls is obtained by differentiating the free
energy $f_N=N\Delta f_N$ with respect to the distance of the walls $L=a(N+1)$:
\begin{equation}
\label{eq:pressm}
p_N=-N\frac{\partial \Delta f_N}{\partial L}=\frac{2N}{N+1}\alpha_N
\frac{k_B^2T^2}{\kappa a^3}.
\end{equation} 
The first and the last Feynman integrals in Eq.~(\ref{constsm}) are the simplest:
\begin{eqnarray}
\label{eq:eight}
\parbox{13mm}{\centerline{
\begin{fmfgraph}(10mm,5mm)
\setval
\fmfforce{0.5w,0.5h}{v}
\fmfforce{0.5w,1h}{vl}
\fmfi{plain}{fullcircle scaled 1h shifted (1/4w,1/2h)}
\fmfi{plain}{fullcircle scaled 1h shifted (3/4w,1/2h)}
\fmfdot{v}
\end{fmfgraph}
}}^{r}&\equiv&\frac{1}{32(N+1)^2}\sum\limits_{n=1}^{N+1}\left[\sum\limits_{m=1}^N
\frac{h_{nn}^m}{\sin(\nu_ma/2)}\right]^2,\\
\label{eq:handle}
\parbox{18mm}{\centerline{
\begin{fmfgraph}(15mm,7mm)
\setval
\fmfforce{0.33w,0.5h}{v1}
\fmfforce{0.66w,0.5h}{v2}
\fmf{plain}{v1,v2}
\fmfi{plain}{reverse fullcircle scaled 0.33w shifted (0.165w,0.5h)}
\fmfi{plain}{fullcircle rotated 180 scaled 0.33w shifted (0.825w,0.5h)}
\fmfdot{v1,v2}
\end{fmfgraph}
}}^{r}&\equiv&\frac{1}{32(N+1)^3}\sum\limits_{n_1,n_2=1}^{N+1}
\sum\limits_{m_1,m_2,\atop m_3=1}^N\frac{h_{n_1n_1}^{m_1}h_{n_1n_2}^{m_2}h_{n_2n_2}^{m_3}}
{\sin(\nu_{m_1}a/2)\sin^2(\nu_{m_2}a/2)\sin(\nu_{m_3}a/2)},
\end{eqnarray}
where we have used the abbreviation $h_{n_1n_2}^m$ defined in Eq.~(\ref{abbrev}).
The evaluation of the second diagram in Eq.~(\ref{constsm})
is much
more involved. The Fourier integrals can be done exactly, except for one, which must be
treated numerically. This calculation is deferred to
Appendix~\ref{app}. The values of the three diagrams are listed in
Table~\ref{mtab2} for various numbers of membranes.
With these numbers, the evaluation of the pressure constants yields the
results given in Table~\ref{mtab3}. Except for $N=1$ and $N\to\infty$, no analytical 
values were found in the literature. We also compare with pressure constants
obtained by Monte-Carlo simulations and find a good agreement~\cite{jkm,GK}.
The values of the Monte-Carlo simulations for $N=3,5,7$
from Ref.~\cite{jkm} show an independence of the number $N$ of membranes. This arises
by the simulation technique, where the free energy of the central membrane
was determined. In contrast to that, we have calculated the pressure constant from the free
energy of the complete system averaged over all membranes. Thus these Monte-Carlo values cannot
be directly compared with ours.

Table~\ref{mtab3} contains also a value $\alpha_\infty$ for an infinite number 
$N\to\infty$ of membranes in the stack. This pressure constant is obtained by
the following extrapolation procedure. We assume that 
the pressure constants determined for $N=12,13,14,15$ are of higher accuracy than
those for lower numbers of membranes. This assumption is justified by comparing
our values for $N=1,3,5$ with the latest Monte-Carlo results~\cite{GK}.    
For $N=1$, the deviation is about $3.4\%$. Considering $N=3$, the deviation reduces
to $1.8\%$ and further to $1.1\%$ for $5$ membranes. Since
the pressure constants is approximated increasingly fast with increasing number $N$ of membranes,
we make the following exponential ansatz for determining the approach to infinite $N$:
\begin{equation}
\label{explaw}
\alpha_N=\alpha_\infty\left[1-\eta\exp\left(-\xi N^\varepsilon \right)\right].
\end{equation} 
The unknown four parameters in this equation are then determined by solving the system of
equations with the pressure constants
$\alpha_{12}$, $\alpha_{13}$, $\alpha_{14}$, and $\alpha_{15}$ listed
in Table~\ref{mtab3}. We obtain $\eta\approx 1.1712$, $\xi\approx 1.6417$, 
$\varepsilon\approx 0.3154$, and thus the limiting pressure constant for an infinite 
stack of membranes 
\begin{equation}
\label{alphainf}
\alpha_\infty\approx 0.1041.
\end{equation}
This value is in very good agreement with the Monte-Carlo result~\cite{GK}
(see the last row of Table~\ref{mtab3}).
It differs by a factor close to $9/4$ from the initial result by Helfrich~\cite{helf1,lca}.
\section{Summary}
We have calculated the pressure constants for a stack of different numbers 
of membranes between two walls 
in excellent agreement with results from Monte-Carlo simulations. 
The requirement that the membranes can not penetrate each other was accounted for 
by introducing a repulsive potential and going to the strong-coupling limit
of hard repulsion. 
We have used the similarity of the membrane system to a stack of strings enclosed by
line-like walls, which is exactly solvable,
to determine the potential parameters in such a way that the two-loop result is exact.
This minimizes the neglected terms in the variational perturbation expansion, when
applying the same potential to membranes.
\section*{Acknowledgements}
We thank Professors W.~Janke and R.~Lipowsky for stimulating discussions 
in this subject. 
One of us (M.B.) is grateful for generous support by the Studienstiftung des deutschen 
Volkes.
\begin{appendix}
\section{Evaluation of the Sunset Diagram}
\label{app}
The second diagram in Eq.~(\ref{constsm}) requires some simplification before the 
numerical calculation. We write the reduced Feynman integral as:
\begin{equation}
\label{sun}
\parbox{9mm}{\centerline{
\begin{fmfgraph}(6mm,6mm)
\setval
\fmfforce{0w,0.5h}{v1}
\fmfforce{1w,0.5h}{v2}
\fmf{plain,left=1}{v1,v2,v1}
\fmf{plain}{v1,v2}
\fmfdot{v1,v2}
\end{fmfgraph}
}}^{r}\equiv\frac{8}{A(N+1)^3}\sum\limits_{n_1,n_2=1}^{N+1}\sum\limits_{m_1,m_2,\atop
m_3=1}^N h_{n_1n_2}^{m_1} h_{n_1n_2}^{m_1} h_{n_1n_2}^{m_1}K_{m_1m_2m_3}
\end{equation}
with the integral
\begin{equation}
\label{intpart}
K_{m_1m_2m_3}=\int d^2x_1 d^2x_2\int\frac{d^2k_1}{(2\pi)^2}\frac{d^2k_2}{(2\pi)^2}
\frac{d^2k_3}{(2\pi)^2}
\frac{e^{-i({\bf k}_1+{\bf k}_2+{\bf k}_3)({\bf x}_1-{\bf x}_2)}}
{({\bf k}_1^4+2\sin^2\nu_{m_1}a)({\bf k}_2^4+2\sin^2\nu_{m_2}a)
({\bf k}_3^4+2\sin^2\nu_{m_3}a)}.
\end{equation}
All integrals 
are easily calculated, except for one. If we introduce
abbreviations
\begin{equation}
\label{mass}
M_l^2=2\sin^2\nu_{m_l}a,\quad l=1,2,3,
\end{equation} 
we find
\begin{equation}
\label{intpart2}
K_{m_1m_2m_3}=\frac{A}{2\pi}\int\limits_0^\infty dk\,\frac{k}{{\bf k}^4+M_3^2}
J({\bf k},M_1^2,M_2^2)
\end{equation}
with
\begin{equation}
\label{J1}
J({\bf k},M_1^2,M_2^2)=\int \frac{d^2p}{(2\pi)^2}\frac{1}{({\bf p}-{\bf k})^4+M_1^2}
\frac{1}{{\bf p}^4+M_2^2}.
\end{equation}
Decomposing the integrand into partial fractions
\begin{eqnarray}
\label{J2}
J({\bf k},M_1^2,M_2^2)&=&-\frac{1}{4M_1M_2}\int \frac{d^2p}{(2\pi)^2}
\left[\frac{1}{({\bf p}-{\bf k})^2+iM_1}-\frac{1}{({\bf p}-{\bf k})^2-iM_1} \right]
\left[\frac{1}{{\bf p}^2+iM_2}-\frac{1}{{\bf p}^2-iM_2} \right]\nonumber\\
&=&-\frac{1}{4M_1M_2}\left[I({\bf k},M_1,M_2)-I({\bf k},M_1,-M_2)
-I({\bf k},-M_1,M_2)+I({\bf k},-M_1,-M_2) \right]
\end{eqnarray}
we are left with integrals of the type
\begin{equation}
\label{I1}
I({\bf k},\gamma_1,\gamma_2)=\int\frac{d^2p}{(2\pi)^2}\frac{1}{({\bf p}-{\bf k})^2+i\gamma_1}
\frac{1}{{\bf p}^2+i\gamma_2},
\end{equation}
where $\gamma_{1,2}=\pm M_{1,2}$ are real numbers.
Employing Feynman's parametrization, these integrals become
\begin{equation}
  \label{eq:feynpar}
  I({\bf k},\gamma_1,\gamma_2)=\frac{1}{4\pi}\int\limits_0^1dx\,
\frac{1}{-x^2k^2+x(k^2+i\gamma_1-i\gamma_2)+i\gamma_2},
\end{equation}
taking the general form
\begin{equation}
  \label{eq:standard}
  \int dx\frac{1}{ax^2+bx+c}=\frac{2}{\sqrt{\Delta}}\arctan{z(x)}
\end{equation}
with
\begin{equation}
  \label{eq:params}
 \Delta=4ac-b^2,\quad z(x)=\frac{b+2ax}{\sqrt{\Delta}},\quad a=-k^2,
\quad b=k^2+i(\gamma_1-\gamma_2),\quad c=i\gamma_2.
\end{equation}
Since $b$ is a complex number, ${\rm Re}\, \arctan z$ is discontinuous, 
if ${\rm Re}\, z$ changes sign and $|{\rm Im}\, z|>1$. Thus 
the right-hand
side of Eq.~(\ref{eq:standard}) is discontinuous at a certain point $x_0$ within the 
interval $[0,1]$. As will be seen subsequently, $J({\bf k},M_1^2,M_2^2)$ 
from Eq.~(\ref{J1}) must be real and thus all
imaginary contributions in the decomposed form~(\ref{J2}) cancel each other. 

We determine the point of discontinuity $x_0$ to obtain the solution of the
integral~(\ref{eq:feynpar}) by investigating the zero of the real part of $z(x)$. 
Decomposing $z(x_0)$ into
real and imaginary part, we obtain
\begin{eqnarray}
\label{eq:repart}
{\rm Re}\, z(x)=|\Delta|^{-1/2}\left[k^2(1-2x)\cos\left(\frac{1}{2}
\arctan \frac{{\rm Re}\,\Delta}{{\rm Im}\,\Delta} \right)+(\gamma_1-\gamma_2)\sin\left(\frac{1}{2}
\arctan \frac{{\rm Re}\,\Delta}{{\rm Im}\,\Delta} \right)\right],\\  
\label{eq:impart}
{\rm Im}\, z(x)=|\Delta|^{-1/2}\left[(\gamma_1-\gamma_2)\cos\left(\frac{1}{2}
\arctan \frac{{\rm Re}\,\Delta}{{\rm Im}\,\Delta} \right)-k^2(1-2x)\sin\left(\frac{1}{2}
\arctan \frac{{\rm Re}\,\Delta}{{\rm Im}\,\Delta} \right)\right],
\end{eqnarray}
where
\begin{equation}
  \label{eq:D}
  {\rm Re}\, \Delta=(\gamma_1-\gamma_2)^2-k^4,\quad {\rm Im}\,\Delta=-2k^2(\gamma_1+\gamma_2).
\end{equation}
Thus, the zero of ${\rm Re}\, z(x)$ is found at
\begin{equation}
  \label{eq:zero}
  x_0=\frac{1}{2}\left\{1+\frac{\gamma_1-\gamma_2}{k^2}\tan\left[\frac{1}{2}\arctan
\frac{2k^2(\gamma_1+\gamma_2)}{k^4-(\gamma_1-\gamma_2)^2} \right] \right\}.
\end{equation}
From the bounds of integration in Eq.~(\ref{eq:feynpar}) follows that we must include 
the discontinuities of (\ref{eq:standard}) for $x_0\in [0,1]$. This 
occurs if $k< |\gamma_1-\gamma_2|$ and ${\rm sign}\, \gamma_1\neq {\rm sign}\, \gamma_2$.
Thus the solution of the integral (\ref{eq:feynpar}) reads
\begin{eqnarray}
  \label{eq:sol}
  I({\bf k},\gamma_1,\gamma_2)=\left\{\begin{array}{ll}   
S({\bf k}, \gamma_1,\gamma_2,x)\Big|_{x=0}^{x=1},\quad\hspace{21.5mm}
\displaystyle{\rm sign}\,\gamma_1={\rm sign}\,\gamma_2 \vee
\displaystyle({\rm sign}\,\gamma_1\neq{\rm sign}\,\gamma_2 \wedge 
k\ge\sqrt{|\gamma_1-\gamma_2|}),\\
\displaystyle\lim_{\varepsilon\to 0}
\left[S({\bf k},\gamma_1,\gamma_2,x)\Big|_{x=0}^{x=x_0-\varepsilon}
+S({\bf k},\gamma_1,\gamma_2,x)\Big|_{x=x_0+\varepsilon}^{x=1}\right],\quad
\begin{array}{l}{\rm sign}\,\gamma_1\neq{\rm sign}\,\gamma_2 
\wedge k<\sqrt{|\gamma_1-\gamma_2|},\end{array}
\end{array}\right.
\end{eqnarray}
where $S({\bf k},\gamma_1,\gamma_2,x)$ is the explicit right-hand side
of Eq.~(\ref{eq:standard}):
\begin{equation}
\label{eq:s}
S({\bf k},\gamma_1,\gamma_2,x)=\frac{1}{2\pi\sqrt{(\gamma_1-\gamma_2)^2-k^4-2ik^2
(\gamma_1+\gamma_2)}}\arctan\frac{k^2(1-2x)+i(\gamma_1-\gamma_2)}
{\sqrt{(\gamma_1-\gamma_2)^2-k^4-2ik^2
(\gamma_1+\gamma_2)}}.
\end{equation}
The function $I({\bf k},\gamma_1,\gamma_2)$ possesses the properties
\begin{eqnarray}
I({\bf k},\gamma_1,-\gamma_2)+I({\bf k},-\gamma_1,\gamma_2)&=&
2{\rm Re}\, I({\bf k},\pm\gamma_1,\mp\gamma_2),\\
I({\bf k},\gamma_1,\gamma_2)+I({\bf k},-\gamma_1,-\gamma_2)&=&
2{\rm Re}\, I({\bf k},\pm\gamma_1,\pm\gamma_2).
\end{eqnarray}
Inserting Eq~(\ref{eq:sol}) into Eq.~(\ref{J2}) and performing
the remaining integral in Eq.~(\ref{intpart2}) together with the
sums in expression~(\ref{sun}) for the sunset diagram can be calculated numerically. The values
are listed for $N=1,\ldots,15$ in the third column of Table~\ref{mtab2}.
\end{appendix}
\newpage
\begin{table}
\caption[]{\label{mtab1} 
Reduced numeric values $W^{{\rm s},r}$ of the two-loop diagrams for the 
free energy for a stack of $N$ strings.
}
\begin{tabular}{cccc}
$N$ & 
$\parbox{12mm}{\centerline{
\begin{fmfgraph}(10mm,5mm)
\setval
\fmfforce{0.5w,0.5h}{v}
\fmfforce{0.5w,1h}{vl}
\fmfi{plain}{fullcircle scaled 1h shifted (1/4w,1/2h)}
\fmfi{plain}{fullcircle scaled 1h shifted (3/4w,1/2h)}
\fmfdot{v}
\end{fmfgraph}
}}^{{\rm s},r}$ &
$\parbox{8mm}{\centerline{
\begin{fmfgraph}(6mm,6mm)
\setval
\fmfforce{0w,0.5h}{v1}
\fmfforce{1w,0.5h}{v2}
\fmf{plain,left=1}{v1,v2,v1}
\fmf{plain}{v1,v2}
\fmfdot{v1,v2}
\end{fmfgraph}
}}^{{\rm s},r}$ &
$\parbox{17mm}{\centerline{
\begin{fmfgraph}(15mm,7mm)
\setval
\fmfforce{0.33w,0.5h}{v1}
\fmfforce{0.66w,0.5h}{v2}
\fmf{plain}{v1,v2}
\fmfi{plain}{reverse fullcircle scaled 0.33w shifted (0.165w,0.5h)}
\fmfi{plain}{fullcircle rotated 180 scaled 0.33w shifted (0.825w,0.5h)}
\fmfdot{v1,v2}
\end{fmfgraph}
}}^{{\rm s},r}$\\ \hline
$1$ & $1/2$ & $0$ & $0$\\
$2$ & $1.288675$ & $0.398717$ & $0.089316$\\
$3$ & $2.100656$ & $0.832299$ & $0.146447$\\
$4$ & $2.915827$ & $1.270787$ & $0.184463$\\
$5$ & $3.730993$ & $1.709326$ & $0.211325$\\
$6$ & $4.545586$ & $2.147034$ & $0.231245$\\
$7$ & $5.359574$ & $2.583849$ & $0.246583$
\end{tabular}
\end{table}
\begin{table}
\caption[]{\label{mtab2} 
Numeric values $W^{r}$ of the reduced two-loop Feynman integrals contributing to the 
pressure constants of a stack of $N$ membranes in Eq.~(\ref{constsm}).
}
\begin{tabular}{cccc}
$N$ & 
$\parbox{12mm}{\centerline{
\begin{fmfgraph}(10mm,5mm)
\setval
\fmfforce{0.5w,0.5h}{v}
\fmfforce{0.5w,1h}{vl}
\fmfi{plain}{fullcircle scaled 1h shifted (1/4w,1/2h)}
\fmfi{plain}{fullcircle scaled 1h shifted (3/4w,1/2h)}
\fmfdot{v}
\end{fmfgraph}
}}^{r}$ &
$\parbox{8mm}{\centerline{
\begin{fmfgraph}(6mm,6mm)
\setval
\fmfforce{0w,0.5h}{v1}
\fmfforce{1w,0.5h}{v2}
\fmf{plain,left=1}{v1,v2,v1}
\fmf{plain}{v1,v2}
\fmfdot{v1,v2}
\end{fmfgraph}
}}^{r}$ &
$\parbox{17mm}{\centerline{
\begin{fmfgraph}(15mm,7mm)
\setval
\fmfforce{0.33w,0.5h}{v1}
\fmfforce{0.66w,0.5h}{v2}
\fmf{plain}{v1,v2}
\fmfi{plain}{reverse fullcircle scaled 0.33w shifted (0.165w,0.5h)}
\fmfi{plain}{fullcircle rotated 180 scaled 0.33w shifted (0.825w,0.5h)}
\fmfdot{v1,v2}
\end{fmfgraph}
}}^{r}$\\ \hline
$1$ & $1/32$ & $0$ & $0$\\
$2$ & $0.080542$ & $0.022446$ & $0.005582$\\
$3$ & $0.131291$ & $0.046992$ & $0.009153$\\
$4$ & $0.182239$ & $0.071866$ & $0.011529$\\
$5$ & $0.233187$ & $0.096762$ & $0.013208$\\
$6$ & $0.284099$ & $0.121619$ & $0.014453$\\
$7$ & $0.334973$ & $0.146428$ & $0.015411$\\
$8$ & $0.385815$ & $0.171195$ & $0.016172$\\
$9$ & $0.436630$ & $0.195925$ & $0.016789$\\
$10$ & $0.487422$ & $0.220624$ & $0.017300$\\
$11$ & $0.538197$ & $0.245300$ & $0.017730$\\
$12$ & $0.588958$ & $0.269954$ & $0.018097$\\
$13$ & $0.639706$ & $0.294592$ & $0.018414$\\
$14$ & $0.690444$ & $0.319215$ & $0.018690$\\
$15$ & $0.741174$ & $0.343827$ & $0.018933$
\end{tabular}
\end{table}
\begin{table}
\caption[]{\label{mtab3} 
Pressure constants $\alpha_N$ for different numbers $N$ of membranes in the stack,
calculated from Eq.~(\ref{constsm}), with the numerical values of the two-loop diagrams
given in Table~\ref{mtab2}. We compare with results from Monte-Carlo simulations
and earlier analytic results. 
}
\begin{tabular}{cccc}
$N$ & $\alpha_N$ & Monte Carlo results & earlier analytic values \\ \hline
$1$ & $\pi^2/128\approx 0.07711$ & $0.060$~\cite{jk}, $0.078\pm 0.001$~\cite{jkm},
$0.0798\pm 0.0003$~\cite{GK}, $0.080$~\cite{jkm} & $\pi^2/128$~\cite{janke,mem},
$0.079715$~\cite{mem}\\ 
$2$ & $0.08669$ & $$ & $$\\
$3$ & $0.09134$ & $0.093\pm 0.004$~\cite{GK}, $0.1002\pm 0.0006$~\cite{jkm} & $$\\
$4$ & $0.09408$ & $$ & $$\\
$5$ & $0.09590$ & $0.0966$~\cite{GK}, $0.1022\pm 0.0006$~\cite{jkm} & $$\\
$6$ & $0.09719$ & $$ & $$\\
$7$ & $0.09815$ & $0.1009\pm 0.0007$~\cite{jkm} & $$\\
$8$ & $0.09890$ & $$ & $$\\
$9$ & $0.09950$ & $$ & $$\\
$10$ & $0.09999$ & $$ & $$\\
$11$ & $0.10039$ & $$ & $$\\
$12$ & $0.10074$ & $$ & $$\\
$13$ & $0.10103$ & $$ & $$\\
$14$ & $0.10129$ & $$ & $$\\
$15$ & $0.10151$ & $$ & $$\\
$\infty$ & $0.10409$ & $0.074$~\cite{jk2}, $0.101\pm 0.002$~\cite{jkm}, $0.106$~\cite{GK} 
& $3\pi^2/128\approx 0.23$~\cite{helf1} \\
\end{tabular}
\end{table}
\end{fmffile}


\begin{thebibliography}{199}
\bibitem{lca}
An analogous estimate for a stack of strings\cite{janke} with Helfrich's method~\cite{helf1}
is also too large by a factor $9/4$.
\end{thebibliography}
\end{document}